\documentclass[prl, 
superscriptaddress,aps, bibliography, twocolumn, reprint, report, showpacs, footnoteinbib]{revtex4-1}

\usepackage{graphicx}
\usepackage{amssymb}   % for math
\usepackage{amsfonts}
\usepackage{amsmath}
\usepackage{dsfont}
\usepackage{natbib} 
\usepackage{dcolumn}   % needed for some tables
\usepackage{bm}        % for math
\usepackage[mathscr]{eucal}
\usepackage[dvipsnames]{xcolor}
\usepackage[colorlinks, linkcolor=OrangeRed,citecolor=RoyalBlue,urlcolor=NavyBlue]{hyperref}
\usepackage[all]{hypcap} % let hyperlinks correctly point to figures rather than their captions; 
\usepackage{mathtools}
\usepackage{wasysym}
\usepackage{graphicx}
\usepackage{amsmath,amssymb,bm}
\usepackage{enumerate}
\usepackage{braket}        % Dirac notation

\usepackage{amsmath}

\newcommand{\eq}[1]{\begin{align}#1\end{align}}

\begin{document}

%%%%%%%%%%%%%%%%%%%%%%%%%%%%%%%%%%%%%%%%%%%%%%%%%
\title{Characterizing time-irreversibility in disordered fermionic systems by the effect of local perturbations}
%%%%%%%%%%%%%%%%%%%%%%%%%%%%%%%%%%%%%%%%%%%%%%%%%
\author{Shreya Vardhan}
\affiliation{Department of Physics, Harvard University, Cambridge MA 02138, USA}
\author{Giuseppe De Tomasi}
\author{Markus Heyl}
\affiliation{Max-Planck-Institut f\"ur Physik komplexer Systeme, N\"othnitzer Stra{\ss}e 38,  01187-Dresden, Germany}
\author{Eric J. Heller}
\affiliation{Department of Physics, Harvard University, Cambridge MA 02138, USA}
\author{Frank Pollmann}
\affiliation{Max-Planck-Institut f\"ur Physik komplexer Systeme, N\"othnitzer Stra{\ss}e 38,  01187-Dresden, Germany}
\affiliation{Technische Universit\"at M\"unchen, 85747 Garching, Germany}
%%%%%%%%%%%%%%%%%%%%%%%%%%%%%%%%%%%%%%%%%%%%%%%%%
%\date{\today}
%\pacs{ } 
%%%%%%%%%%%%%%%%%%%%%%%%%%%%%%%%%%%%%%%%%%%%%%%%%%%%%%%%%5
\begin{abstract}
We study the effects of local perturbations on the dynamics of disordered fermionic systems in order to characterize time-irreversibility. We focus on three different systems, the non-interacting Anderson and Aubry-Andr\'e-Harper (AAH-) models, and the interacting spinless disordered t-V  chain. First, we consider the effect on the full many-body wave-functions by measuring the Loschmidt echo (LE). 
We show that in the extended/ergodic phase the LE decays exponentially fast with time, 
while in the localized phase the decay is algebraic. We demonstrate that the exponent of the decay of the LE in the localized phase diverges proportionally to the single-particle localization length as we approach the metal-insulator transition in the AAH model. 
Second, we probe different phases of disordered systems by studying the time expectation value of local observables evolved with two Hamiltonians that differ by a spatially local perturbation. Remarkably, we find 
that many-body localized systems could lose memory of the initial state in the long-time limit, in contrast to the non-interacting localized phase where some memory is always preserved.
\end{abstract}
%%%%%%%%%%%%%%%%%%%%%%%%%%%%%%%%%%%%%%%%%%%%%%%%%%%%%%%%%5
\maketitle
%%%%%%%%%%%%%%%%%%%%%%%%%%%%%%%%%%%%%%%%%%%%%%%%%%%%%%%%%%%%%%%%%%%%%%5common
{\it Introduction}---The second law of thermodynamics imposes strong constraints on the time-reversibility of non-adiabatic processes between thermodynamic states. However, the applicability of the second law requires, in general, ergodicity, which is absent for closed many-body localized (MBL) systems~\cite{Basko06,Gor05, Pr08, Pal10, Ros11, De13, Lev14, FP14, Be15, luitz15, Rajeev15, Rajeev16, Burr15}. This has recently also been shown experimentally~\cite{Schreiber15, Smith16, Choi16, Bordia16, Lus16}. Consequently, after a non-adiabatic process, such MBL systems do not inherit a thermodynamic description, leading to a major question: to what extent does this breaking of ergodicity influence time-reversibility?

In this work, we probe time-reversibility of closed quantum many-body systems with disorder. Specifically, we study the sensitivity of the dynamics due to weak local perturbations for a wide range of fermionic systems including the Aubry-Andr\'e-Harper (AAH) as well as the Anderson model with and without interactions. We characterize time-reversibility by noticing that the faster the departure of the  perturbed and unperturbed trajectories, the stronger is the sensitivity of quantum motion and therefore the stronger time-irreversibility. In this work, the quantification of the distance of the two time-evolved systems is based on two complementary measures: First, we study the sensitivity in terms of the Loschmidt echo (LE)~\cite{Peres84, Gorin06}, which quantifies the deviation on the basis of the full quantum many-body wave-function. Second, we introduce a quantity which measures the closeness of only local properties instead of global wave functions, by measuring the local densities, representing a less strict measure as compared to the LE. We find numerical evidence corroborated by analytical arguments that the various distinct phases of our fermionic models can be detected and characterized by studying the long-time dynamics of these measures. 
Our predictions can be tested experimentally because both of the studied quantities are, in principle, experimentally accessible for ultracold atoms in optical lattices and trapped ions where signatures of MBL have been already observed recently~\cite{Schreiber15, Smith16, Choi16, Bordia16, Lus16}. In systems of ultracold atoms, local densities can be measured with the use of quantum gas microscopes~\cite{Bakr2009,Sherson2010} and LE by interferometric techniques~\cite{Knap12, Daley12,Pich13}. The local control of trapped ions provides direct access to local densities and LE have been already measured in recent experiments~\cite{Martinez2016,Jur16}.

{\it Models and methods}---We study the Hamiltonian
\begin{equation*} 
 \hat{\mathcal{H}}:=-\frac{t_{h}}{2}\sum_{x=-\frac{L}{2}}^{\frac{L}{2} -2}\hat{c}^\dagger_{x}\hat{c}_{x+1}+h.c.+\sum_{x=-\frac{L}{2}}^{\frac{L}{2} -1}h_x\hat{\tilde{n}}_x+V\sum_{x=-\frac{L}{2}}^{\frac{L}{2} -2}\hat{\tilde{n}}_x \hat{\tilde{n}}_{x+1}
\label{Eq:Ham}
\end{equation*}
\begin{figure}
 \includegraphics[width=1.\columnwidth]{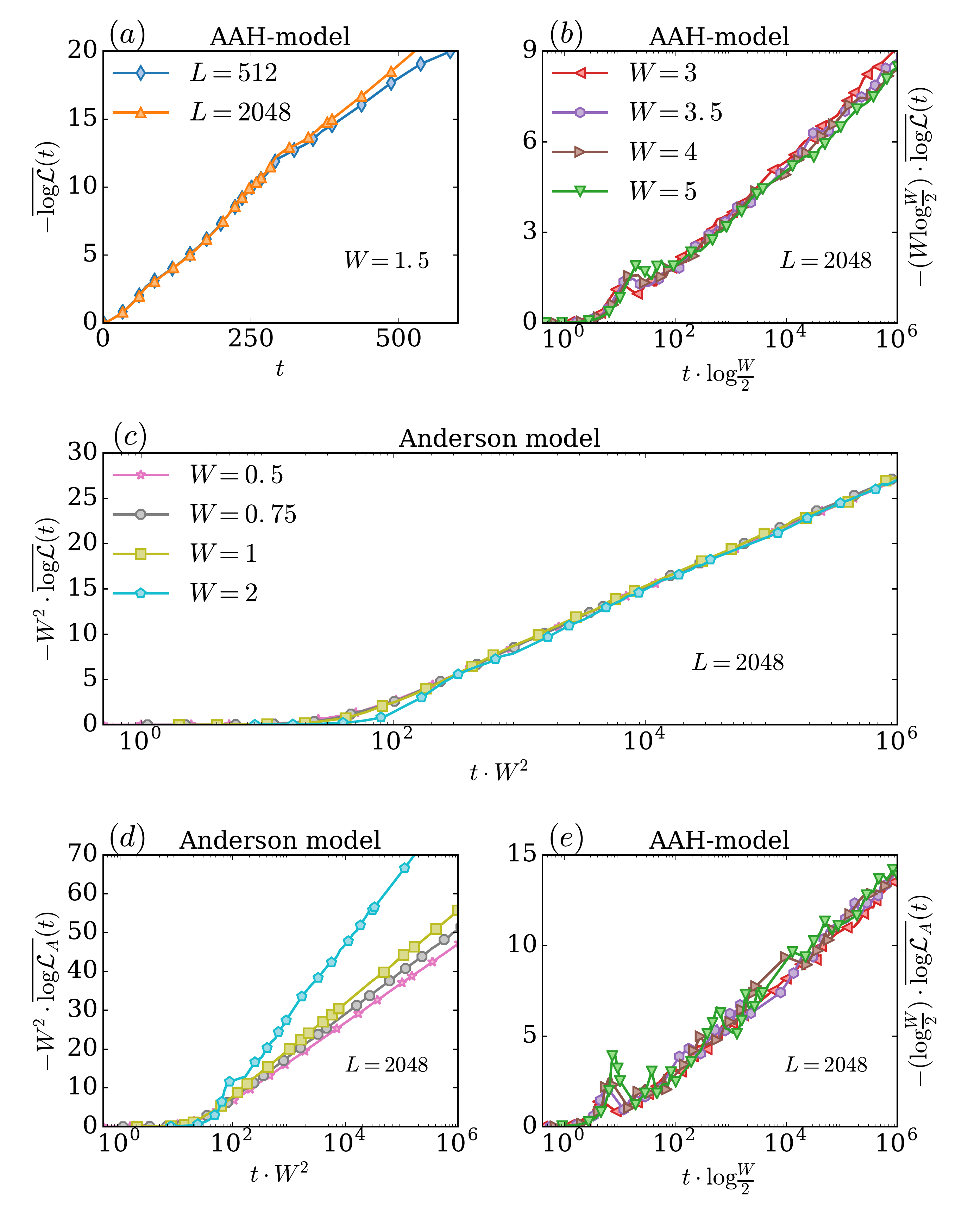}
 \caption{(a),(b): Behavior of $-\overline{\log{\mathcal{L}}}(t)$ for the AAH model in the extended phase ($W=1.5$)  $(\mathcal{L}(t)\sim e^{-\Gamma t})$ and in the localized phase for several values of $W$ $(\mathcal{L}(t)\sim t^{-\beta})$. In the localized phase $t$ and $\mathcal{L}(t)$ have been properly rescaled to underline 
the time scale on which the decays starts and the behavior of the exponent of the algebraic decay $\beta$. (c): $-\overline{\log{\mathcal{L}}}(t)$ for the Anderson model for several values of $W$; here also a rescaling has been done on $t$ and $\mathcal{L}(t)$. (d),(e): Panels show the approximate formula $\mathcal{L}_A(t)$ for the two non-interacting models and for the same 
values of $W$.}
\label{fig:A_echo}
\end{figure}where $\hat{c}_x^\dagger~(\hat{c}_x)$ is the fermionic creation (annihilation) operator at
site $x$ and $\hat{\tilde{n}}_x$=$\hat{n}_x-\frac{1}{2}$ with $\hat{n}_x$=$\hat{c}_x^\dagger\hat{c}_x$, $L$ the system size and $N$=$\frac{L}{2}$ the number of fermions. 

We consider three different cases: (i) The non-interacting Aubry-Andr\'e-Harper (AAH-) model, obtained from  $\hat{\mathcal{H}}$ with $V$=$0$, $t_{h}$=$2$ and  
$h_x$=$W\cos(2\pi x\phi+\alpha)$ where $\phi$=$\frac{1+\sqrt{5}}{2}$; $\alpha$ is a random phase uniformly distributed in $[0,2\pi]$. The AAH model has a metal-insulator transition at $W_c$=$2$ (extended phase for $W\le W_c$ and localized phase for $W>W_c$). The localization length close to the 
transition diverges as $\xi_{\text{loc}}\sim \log^{-1}\frac{W}{2} $ ~\cite{Au80}. (ii) The non-interacting Anderson model~\cite{Anderson58}, given by $V$=$0$, $t_{h}$=$1$ and  $\{h_x\}$ independent random variables uniformly distributed in $[-W,W]$. In the Anderson model, all the single-particle
eigenstates are exponentially localized and $\xi_{\text{loc}} \sim W^{-2}$ in the weak disorder regime~\cite{Derr84}. 
(iii) The spinless disordered t-V  chain, obtained from the Anderson model by turning on the interaction with $V$=$1$. This t-V  chain is believed to have a many-body localization (MBL) transition at a critical disorder strength $W_c \approx 3.5$ \cite{luitz15, Pal10, Be15, GTDM} (extended/ergodic for $W < W_c$ and localized for $W>W_c$) at infinite temperature.

To study a spatially local perturbation of the Hamiltonian $\hat{\mathcal{H}}$, we define 
\begin{equation}
 \hat{\mathcal{H}}_{\epsilon}  := \hat{\mathcal{H}} + 2\epsilon \hat{n}_0,
\end{equation}
with $\epsilon >0$.
A central object studied in this work is the LE~\cite{Poll10,Dora16,Dora13,Lev98,Cu03,Jac01}, which in related forms has already been studied in disordered systems~\cite{Adamov03,Body09,Roy15,Knap14,Deng15}
\begin{equation}
 \mathcal{L} (t)  :=  | \langle \psi | e^{it   \hat{\mathcal{H}}}  e^{-it \hat{\mathcal{H}}_\epsilon } |\psi \rangle |^2.
 \label{Eq:L_echo}
\end{equation}
To understand how states deviate in their local properties if evolved with $\mathcal{H}$ and $\mathcal{H}_\epsilon$, we study the difference of the local density profile (DLDP)~\cite{Khemani2015,Deng15}, defining 
\begin{equation}
\mathcal{D}(t)  := \sum_x |\delta \rho(x,t)|
\end{equation}
with
\begin{equation}
 \delta \rho(x,t)  :=   \langle \psi | e^{it\hat{\mathcal{H}}} \hat{n}_x  e^{-it\hat{\mathcal{H}}} | \psi \rangle -  \langle \psi | e^{it\hat{\mathcal{H}}_\epsilon} \hat{n}_x  e^{-it\hat{\mathcal{H}}_\epsilon} | \psi \rangle,
\end{equation}
Moreover, we are interested in the long-time behavior of $\mathcal{D}(t)$, which quantifies the long-time relative temporal fluctuations 
\begin{equation}
\mathcal{D}_\infty := \lim_{T\rightarrow \infty} \frac{1}{T}\int_{0}^{T}ds \mathcal{D}(s).
\end{equation}
For the initial state $|\psi \rangle$, we choose a product state in the occupation basis $\big (\prod_{s=1}^N c^\dagger_{2s} |0\rangle \big )$ (Neel-state), which is easy to realize in experiments ~\cite{Schreiber15}. The strength of the perturbation $\epsilon$ is set equal to $0.1$, so $\epsilon < \{t_h,W,V\}$. 
The average over disorder is indicated with an overline \footnote{See supplemental material for additional data on the dependence of the initial state and on the perturbation strength $\epsilon$.}, e.g. $\overline{\mathcal{D}}(t)$.

{\it Non-interacting models}---
In this section, we study the LE and the DLDP for the AAH- and Anderson-models. We compute the LE for these models using a free fermion technique~\cite{Pes09}, which permits us to inspect large system sizes for long times.
Figure \ref{fig:A_echo} (a-c) shows the LE in the two phases of the AAH- and in the Anderson model. In the extended phase of the AAH-model (W = 1.5), the LE decays exponentially as $\mathcal{L} (t) \sim e^{-\Gamma t}$, revealing the strong effect of local small perturbations. In the localized  phase for both models (AAH- and Anderson-model), the LE decays algebraically in time as $\mathcal{L} (t) \sim t^{-\beta}$. Note that in both phases, the long time saturation value is exponentially small in system size, i.e. $\mathcal{L} (t\rightarrow  \infty) \sim e^{-\eta L}$. Still the two phases can be distinguished through the decay of the LE as a function of time.

For the localized phase, Fig.~\ref{fig:A_echo} (b-c) also shows the relation between the exponent $\beta$  and the microscopic parameter of the Hamiltonian ($W$), with a good collapse of the curves. For the Anderson model, we observe $\beta \propto W^{-2}$, indicating that
 $\beta$ is proportional to $\xi_{\text{loc}}$ at least in the weak disorder limit. For the AAH-model, we find the scaling $\beta \propto (W\log{\frac{W}{2}})^{-1} $. Thus, $\beta$ is again proportional to the localization length $\xi_{\text{loc}}$ on approaching the metal-insulator transition to leading order. The rescaled time in the LE deserves particular attention: the time scale for the onset of the algebraic decay is proportional to the localization length, which on approaching the metal-insulator transition shifts to infinity. 
 
We now present an analytical argument supporting the algebraic decay of the LE in the localized phase. 
In the Lehmann representation the LE reads
\begin{equation}
 \mathcal{L} (t)  =  \left | \sum_{n,m}  \langle \psi | n \rangle \langle n | m_{\epsilon}\rangle \langle m_{\epsilon} |\psi \rangle e^{-it(E_n-E_m^{(\epsilon) })}  \right |^2,
 \label{Eq:L_loc}
\end{equation}
where $E_n$ $(| n \rangle)$ and $E_m^{(\epsilon)}$ $(|m_{\epsilon}\rangle)$ are the eigenvalues (eigenvectors) of $\hat{\mathcal{H}}$ and $\hat{\mathcal{H}_\epsilon}$, respectively.
The simple picture is that in the localized phase, the local perturbation causes an exponentially weak dephasing of the energies of the unperturbed Hamiltonian with respect to the perturbed one, inducing the decay of LE. The following approximations, which are equivalent to a first order expansion in $\epsilon$~\cite{Cer03}, permit us to estimate the behavior of the LE and relate the power-law exponent $\beta$ to the localization length. We confirmed this relation close to the metal-insulator transition with exact numerics. First, we assume that the behavior of the LE is not affected by the choice of the initial product state. Our second approximation is that the perturbation affects only the eigenenergies but not the eigenstates. %This "diagonal approximation" has been applied in the context of quantum chaos for one-particle systems ~\cite{Cer03}. 

The first approximation allows us to replace the overlap with the initial state $|\psi \rangle$ in Eq.~\eqref{Eq:L_loc} with a normalized trace over the entire Hilbert space. %( $\left |\langle \psi | n \rangle\right |^2\rightarrow \frac{1}{2^L}$ [IF THIS IS NOT OBVIOUS]), 
The second approximation implies $\langle n | m_{\epsilon}\rangle = \delta_{n,m}$.
%and evaluating the energy difference $E_n-E_n^{(\epsilon) }$ using first order perturbation in $\epsilon$, $E_n-E_n^{(\epsilon)}\approx\epsilon\langle n|\hat{n}_0|n \rangle = \epsilon \sum_{j=1}^{L} a_j^{(n)} |\phi_j(0)|^2$,
%where $\{\phi_j(0)\}$ are the single particle wave-functions evaluated in the center of the chain and $a_j^{(n)}$ takes values $\{1,0\}$ regarding if the single particle mode label with $j$ is occupied or 
%not in the eigenstate $|n\rangle$. The last result can be resumed giving
Finally, evaluating the energy difference $E_n-E_n^{(\epsilon) }$ using first-order perturbation theory in $\epsilon$, $E_n-E_n^{(\epsilon)}\approx\epsilon\langle n|\hat{n}_0|n \rangle$, we can express the result in a closed form
\begin{equation}
  \mathcal{L}_A (t) = \prod_{j=1}^L \cos^2 \left ( \epsilon |\phi_j(0)|^2 t\right ),
 \label{Eq:approx_formula}
  \end{equation}
where $\{\phi_j(0)\}$ are the single-particle wave-functions evaluated at the center of the chain. The subscript $A$ underlines that this is an approximate formula.
Since all single particle eigenstates are exponentially localized, after an appropriate relabeling of the index $j$, we assume  that $|\phi_j(0)|^2 \sim \frac{e^{-\frac{j}{\xi}}}{\xi}$. Thus, the only factors that contribute significantly are the ones where $\epsilon |\phi_j(0)|^2 t \approx 1$
\begin{equation}
  \mathcal{L}_A (t) \approx \prod_{j=1}^{\xi \log{ \frac{\epsilon t}{\xi}}} \cos^2 \left ( \epsilon |\phi_j(0)|^2 t\right ) \sim \left (\frac{\epsilon t}{\xi}\right )^{-c\xi},
  \label{Eq:time_scale}
  \end{equation}
 with $c>0$. %Computing numerically the approximated formula $\mathcal{L}_A(t)$, 
 The last row of Fig.~\ref{fig:A_echo} (c,d) shows the algebraic decay with time of the LE from Eq.~\eqref{Eq:approx_formula} as $\mathcal{L}_A (t)\sim t^{-\beta_A}$ for the two models and several values of $W$.
Surprisingly, despite being a perturbative expansion in $\epsilon$, $\mathcal{L}_A(t)$ reproduces the algebraic decay of the LE also for long times. The exponents $\beta_A$ and $\beta$ have the same dependence on the microscopic parameter $W$ in the vicinity of the critical point, namely 
$\beta,\beta_A\sim W^{-2}$ as $W\rightarrow 0$ for the Anderson-model and 
$\beta,\beta_A\sim\log^{-1}\frac{W}{2}$ as $W\rightarrow 2$ for the AAH-model. Indeed, as shown in Fig.~\ref{fig:A_echo} (d-e), $\beta_A$ is proportional to the localization length $(\beta_A \propto \xi_{\text{loc}})$. For the Anderson model, the deviation with increasing disorder strength $W$ is just a sign that the perturbative expansion for $\xi_{\text{loc}}$ is breaking down.
Moreover, the approximate formula Eq.~\eqref{Eq:time_scale} describes well the rescaling of time, given by $t\rightarrow \frac{\epsilon t}{\xi}$.
 \begin{figure}
 \includegraphics[width=1.\columnwidth]{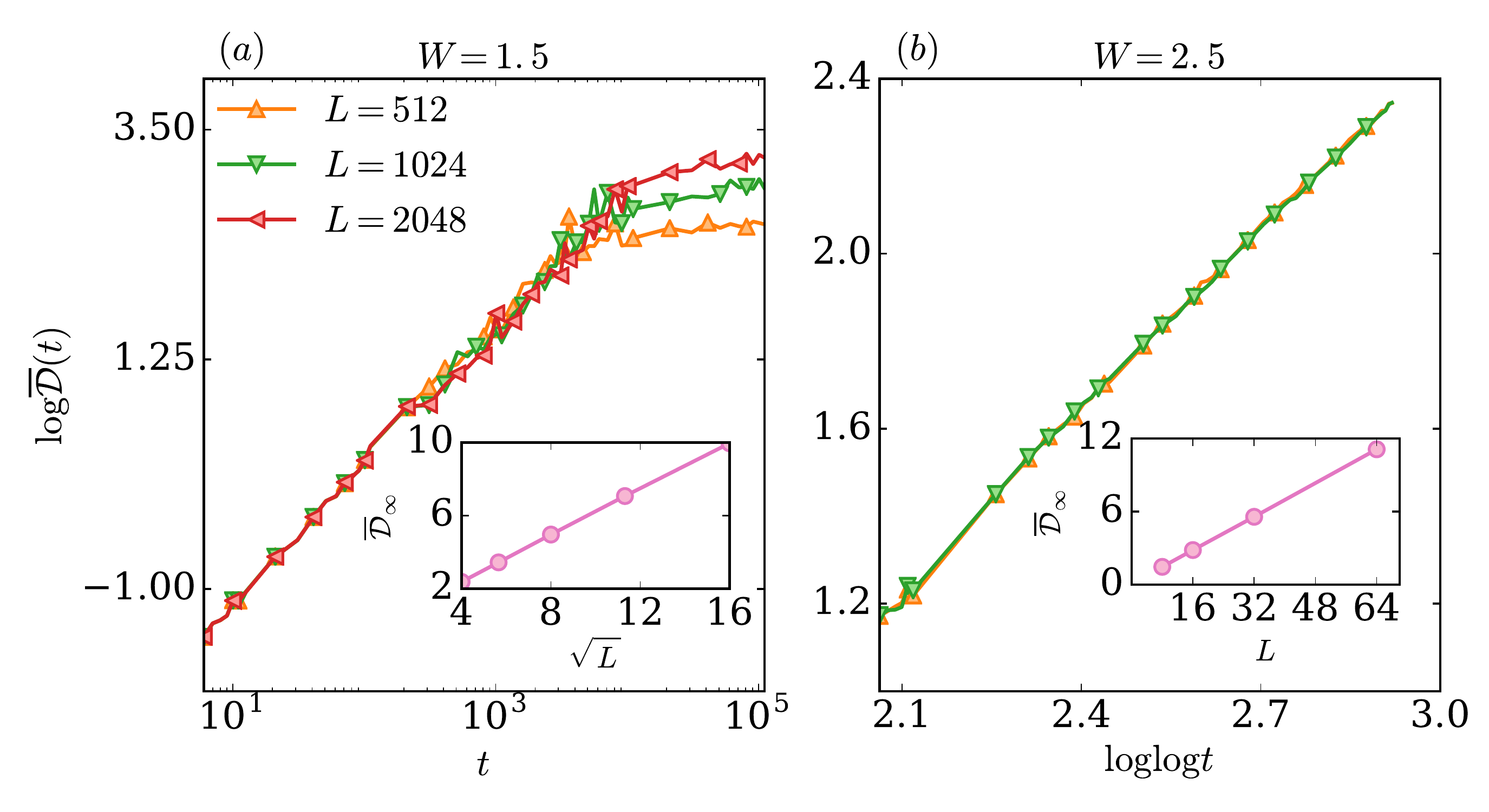}
 \caption{$\overline{D}(t)$ for the AAH model in the two phases for different $L$. (a): $W=1.5$, $\overline{D}(t)\sim t^\alpha$ while its inset shows $\overline{\mathcal{D}}_\infty$ as a function of $L$ ($\overline{\mathcal{D}}_\infty\sim \sqrt{L})$. (b): $W=2.5$, $\overline{\mathcal{D}}(t)\sim \log^\alpha t$, its inset shows $\overline{\mathcal{D}}_\infty\sim L$.}
 \label{fig:D_A}
\end{figure} 

Next, we probe the effect of local perturbation on the dynamics of local observables by studying  $\overline{\mathcal{D}}(t)$. Figure~\ref{fig:D_A} shows $\overline{\mathcal{D}}(t)$ for two different values of $W$ for the
 AAH-model. In the extended phase with $W=1.5$, $\overline{\mathcal{D}}(t)$ shows an algebraic growth with time, $\overline{\mathcal{D}}(t)\sim t^\alpha$, $\alpha \approx 0.6$ for $W=1.5$. The saturation point in time of $\overline{\mathcal{D}}(t)$ is consistent with the scale $\sqrt{L}$ (inset, Fig.~\ref{fig:D_A} (a)) with system size, indicating that in the long time limit the average over index sites of the DLDP ($\frac{\overline{\mathcal{D}}_\infty}{L}$) relaxes  algebraically with system size \footnote{In the supplemental material we provide an analytical argument of a lower bound of the scaling with $L$ for  $\overline{\mathcal{D}}_\infty$}. 
 %The dependence on $L$ is a direct consequence of the fact that the system is non-interacting and extended: the Slater determinant structure is preserved during the time evolution, but at long times, the relative phases associated with different energies become approximately random. So the expectation values  of $\langle \psi | e^{it\mathcal{H}} \hat{n}_x  e^{-it\mathcal{H}} | \psi \rangle$ ($\langle \psi | e^{it\mathcal{H}_\epsilon} \hat{n}_x  e^{-it\mathcal{H}_\epsilon} | \psi \rangle$) 
 %can be described by expectation values over a random Slater determinant $(|\delta \rho(x,t\rightarrow \infty) | \sim \frac{1}{\sqrt{L}})$.
 
 In the localized phase, $\overline{\mathcal{D}}(t)$ has a log-like slow growth, $\overline{\mathcal{D}}(t) \sim \log^\alpha {t}$ with $\alpha \approx 1.3$ for $W=2.5$, so the effect of local perturbations on the dynamics is exponentially slow in time. Moreover, $\overline{\mathcal{D}}_\infty \sim L$ (inset, Fig.~\ref{fig:D_A} (b)), so that the relaxation of $\frac{\overline{\mathcal{D}}_\infty}{L}$ never takes place. 
 %We can use a statistical approach to understand the saturation value of $\overline{\mathcal{D}}_\infty$.
 %The initial state is a Slater determinant of delta-functions placed at different sites. As the model is localized %and non-interacting, the final state at long times can be approximated as a Slater determinant of finite-width %box functions displaced randomly along the chain. 
 %Replacing the expectation values with an expectation over this Slater determinant, we find %$|\delta\rho(x,t\rightarrow \infty)| \sim \mathcal{O}(1)$ \cite{Mask14}. 
  \begin{figure*}[tb]
 \includegraphics[width=1.\textwidth]{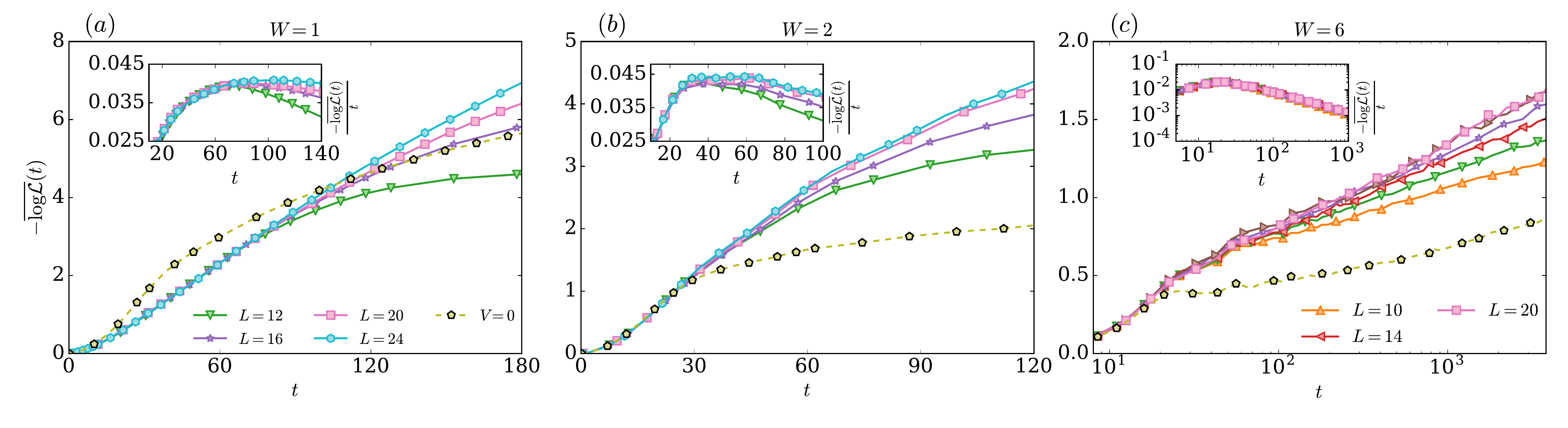}
 \caption{The panels show $-\overline{\log{\mathcal{L}}}(t)$ for different values of disorder strength $W$. (a): The system is in the ergodic phase $W=1$ and LE decays at least exponentially fast with time. (b): An intermediate disorder strength $W=2$,
  $\frac{-\overline{\log{\mathcal{L}}}(t)}{t}$ (inset) forms a plateau with time which is enlarging with system size, showing that the range of times for which LE decays exponentially fast is expanding.
  (c): The system is in the localized phase $W=6$ and LE decays algebraically with time. 
 We also show the LE for the non-interacting case $(V=0)$ for the largest system size in each panel ($L=24$ for $W=1,2$ and $L=20$ for $W=6$).}
 \label{fig:L_Echo_MBL}
 \end{figure*}
 \begin{figure}
 \includegraphics[width=1.\columnwidth]{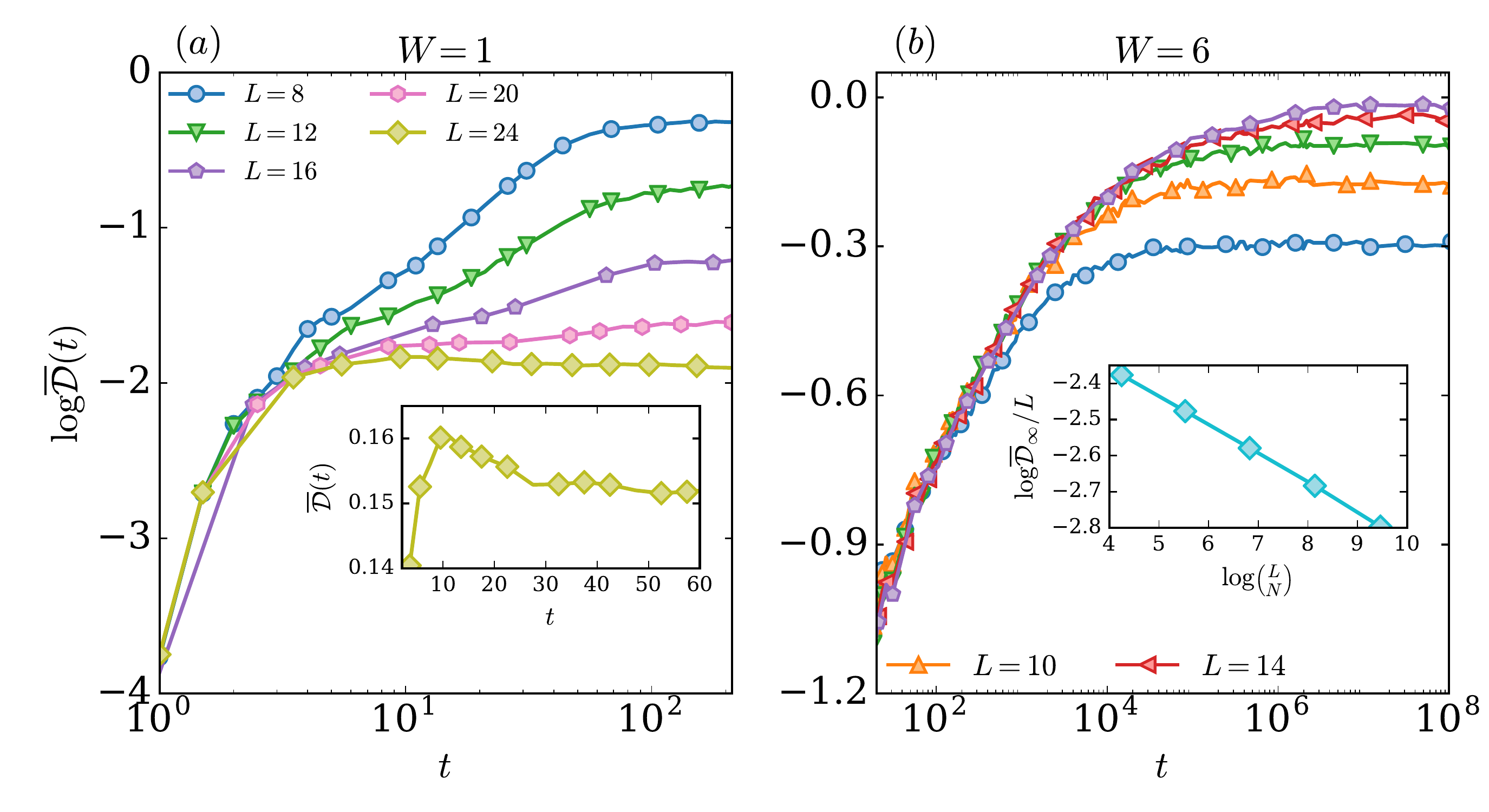}
 \caption{$\overline{\mathcal{D}}(t)$ for the spinless disorder t-V  chain for different $L$ and two values of $W$. (a) $W=1$, the inset shows $\overline{\mathcal{D}}(t)$ for $L=24$ to underline its non-monotonic dependence on $t$. (b) $W=6$, the inset shows $\frac{\mathcal{D_\infty}}{L}$ as function of $L$, it decays exponentially fast with $L$, $\frac{\overline{\mathcal{D}}_\infty}{L} \sim \binom{L}{N}^{-\gamma}$.}
 \label{fig:D_A_MBL}
\end{figure}

{\it Spinless t-V  chain}---Having shown that the LE captures the salient features of the metal-insulator transition in the AAH model,  we now study $\mathcal{L}(t)$ for the interacting spinless t-V chain that has an MBL transition. We perform the time evolution  
using full diagonalization for small systems size $L\le 16$, and using the Chebyshev integration technique~\cite{Ker06} for larger $L$ $(18\le L \le 24)$.
Figure \ref{fig:L_Echo_MBL} (a-c) shows the behavior of LE for the interacting model for different values of disorder strength $W$. The enhanced decay compared with the non-interacting problem is also shown in Fig.~\ref{fig:L_Echo_MBL}). Nevertheless, in the localized phase, the LE still decays algebraically as in the localized phase of the non-interacting
models. For $W=6$ the function $-\frac{\overline{\log{\mathcal{L}}}(t)}{t}$ (inset Fig.~\ref{fig:L_Echo_MBL} (c)) does not present any systematic dependence on system size, indicating that the algebraic decay could be the asymptotic  thermodynamic behavior. In the ergodic phase with $W=1$, LE decays at least exponentially with time, and the function $-\frac{\overline{\log{\mathcal{L}}}(t)}{t}$ does not decay for times in which the decay of LE is not affected by finite size effects (inset Fig.~\ref{fig:L_Echo_MBL} (a)). Figure~\ref{fig:L_Echo_MBL} (b) also shows an intermediate disorder value $W=2$, at which the function $-\frac{\overline{\log{\mathcal{L}}}(t)}{t}$ develops a plateau with respect to $t$, like in the extended phase, after which a slower decay sets in. This plateau is enlarging with increasing system size, which may indicate that 
in the thermodynamic limit ergodicity will be completely restored and the LE will decay exponentially with $t$.

We now study the effects of perturbations in the dynamics of local observables by studying DLDP.
Figure \ref{fig:D_A_MBL} shows $\overline{\mathcal{D}}(t)$ in the interacting model for two values of $W$. We give evidence that the behavior of $\overline{\mathcal{D}}(t)$ in the ergodic phase for long time is drastically different from the non-interacting case: 
 $\overline{\mathcal{D}}(t)$ is not a monotonic function of $t$ (inset, Fig.~\ref{fig:D_A_MBL} (a)). For short times, $\overline{\mathcal{D}}(t)$ grows to a maximum value from which it starts to decay to a finite $L$-dependent value. The non-monotonic behavior is intimately connected with the thermalization of the system. Indeed, the long time expectation values of local observables for thermal systems at infinite temperature should be unchanged if the system is locally perturbed. 
 %Since the system is ergodic, by the diffusion and scattering of the particles, the memory of the initial state with well-defined particle positions is quickly lost, leading to equilibration between all the degrees of freedom. 
 The average time in which the decay of $\overline{\mathcal{D}}(t)$ starts defines a time scale $\tau$; this is roughly the time at which $\overline{\mathcal{D}} (t)$ changes concavity and starts to decrease. 
For times much larger than $\tau$, the expectation value of a local observable is given by the expectation value over a many-body random state (ETH at infinite temperature), so that $|\delta \rho(x,t\gg \tau)| \sim \binom{L}{N}^{-\gamma}\sim e^{-(\gamma\log{2})L}$. 

In the localized phase, the finite size-effects become more important, and for smaller system sizes it could seem that $\overline{\mathcal{D}}(t)$ \footnote{See supplemental material for additional data for $\mathcal{D}(t)$ in the ergodic and MBL phase.} has an unbounded slow growth similar to the localized phases for the non-interacting models. However, a careful analysis shows that the saturation value is merely an exponential decay such as in the extended phase, consisted with $\overline{\mathcal{D}}_\infty \sim L \binom{L}{N}^{-\gamma}$ (inset, Fig.~\ref{fig:D_A_MBL} (b)). Compared with the ergodic phase, in the localized phase the exponent $\gamma$ is small, so that for the considered system sizes, the behavior of $\overline{\mathcal{D}}_\infty$ is dominated by the linear prefactor $L$. In the thermodynamic 
limit we expect that the final shape will be similar to the one in the ergodic phase, so that $\overline{\mathcal{D}}(t)$ will eventually also decay with time at long times. Note that the time scale at which this decay will take place is extremely large; 
the limitation on system size does not allow us to estimate
an upper bound of the time scale $\tau$, which leaves open the possibility that $\tau$ might shift to infinity with increasing $L$. The behavior of $\overline{\mathcal{D}}_\infty$ in the localized phase is reminiscent of the long time ``volume-law" saturation of the entanglement entropy $\mathcal{S}(t)$ after a quantum quench. The distinction between the ergodic and the MBL phase lies only in the numerical value of the prefactor in front of the saturation value of $S(t)$ ~\cite{Bar12,Aba13}, while the scaling with $L$ is the same in both phases (volume law). 
%Indeed, the long-time saturation of $\mathcal{S}(t)$ in the localized phase shows that 
%the resulting state after quantum quench from a product state is a non-thermal but still volume law entanglement state ~\cite{Bar12,Aba13}, while in the ergodic phase the saturation values is thermal and volume law, so what does distinguish the two phases is the prefactor 

{\it Conclusion}---In this work, we probed the effects of local perturbations on the dynamics of several disordered systems by studying the Loschmidt echo (LE) and the difference of the local density profile (DLDP). First, with a combination of analytical arguments and exact numerical simulations, we showed that the LE in the localized phase decays algebraically in time. Furthermore, we found, for the non-interacting models, that the exponent of the algebraic decay is proportional to the single-particle localization length, which diverges at the metal-insulator transition.  In the extended phase, the LE decays exponentially fast with time. The faster exponential decay in the extended phase compared with the algebraic decay in the localized phase implies that time-irreversibility is more strongly manifested in the extended phase than in the localized phase, at least for local perturbations.  Second, we studied the DLDP for the same models, and we found that the long-time behavior saturates algebraically with system size in the extended phase of the Aubry-Andr\'e-Harper model, 
while it never relaxes for the non-interacting localized phase. For the DLDP in the spinless disordered t-V chain, the relaxation is exponential in system size in both phases: in the ergodic phase this is due to thermalization, while in the MBL phase it could be due to the interaction-induced dephasing mechanism which is also explains the long-time saturation values of the entanglement entropy after a quantum quench. The study of the change in the expectation values of local observables when the system is perturbed, gives a different perspective  concerning time-irreversibility as opposed to the LE. Indeed, the long-time expectation value of local observables in a thermal system at infinite temperature should be unchanged if the system is locally perturbed. We give numerical evidence that this also happens in the MBL phase. 
\begin{acknowledgments}
{\it Note}---In completing the manuscript we have become aware of related works on LE in the MBL phase~\cite{Serb17,Hamma17}.
\end{acknowledgments}

\begin{acknowledgments}
{\it Acknowledgments}---We thank J.H. Bardarson, S. Bera, L. Bucciantini, A. Burin, T. Grover, J.-M. Stephan, S. Roy, S. Tomsovic for several illuminating discussions.  This work was partially supported by DFG Research Unit FOR 1807 through grants no. PO 1370/2-1 and by the Deutsche Forschungsgemeinschaft via the Gottfried Wilhelm Leibniz Prize programme. This research was supported in part by the National Science Foundation under Grant No. NSF PHY-1125915. 
\end{acknowledgments}
%%%%%%%%%%%%%%%%%%%%%%%%%%%%%%%%%%%%%%%%%%%%%%%%%%%%%%%%%5
\bibliography{L_echolBIB}
%%%%%%%%%%%%%%%%%%%%%%%%%%%%%%%%%%%%%%%%%%%%%%%%%%%%%%%%%5
\newpage 
\clearpage
\subsection{Supplemental material to Characterizing time-irreversibility in disordered fermionic systems by the effect of local perturbations}
\subsection{Dependence of $\mathcal{L}(t)$ on the initial state and $\epsilon$} 
\begin{figure}
 \includegraphics[width=1.\columnwidth]{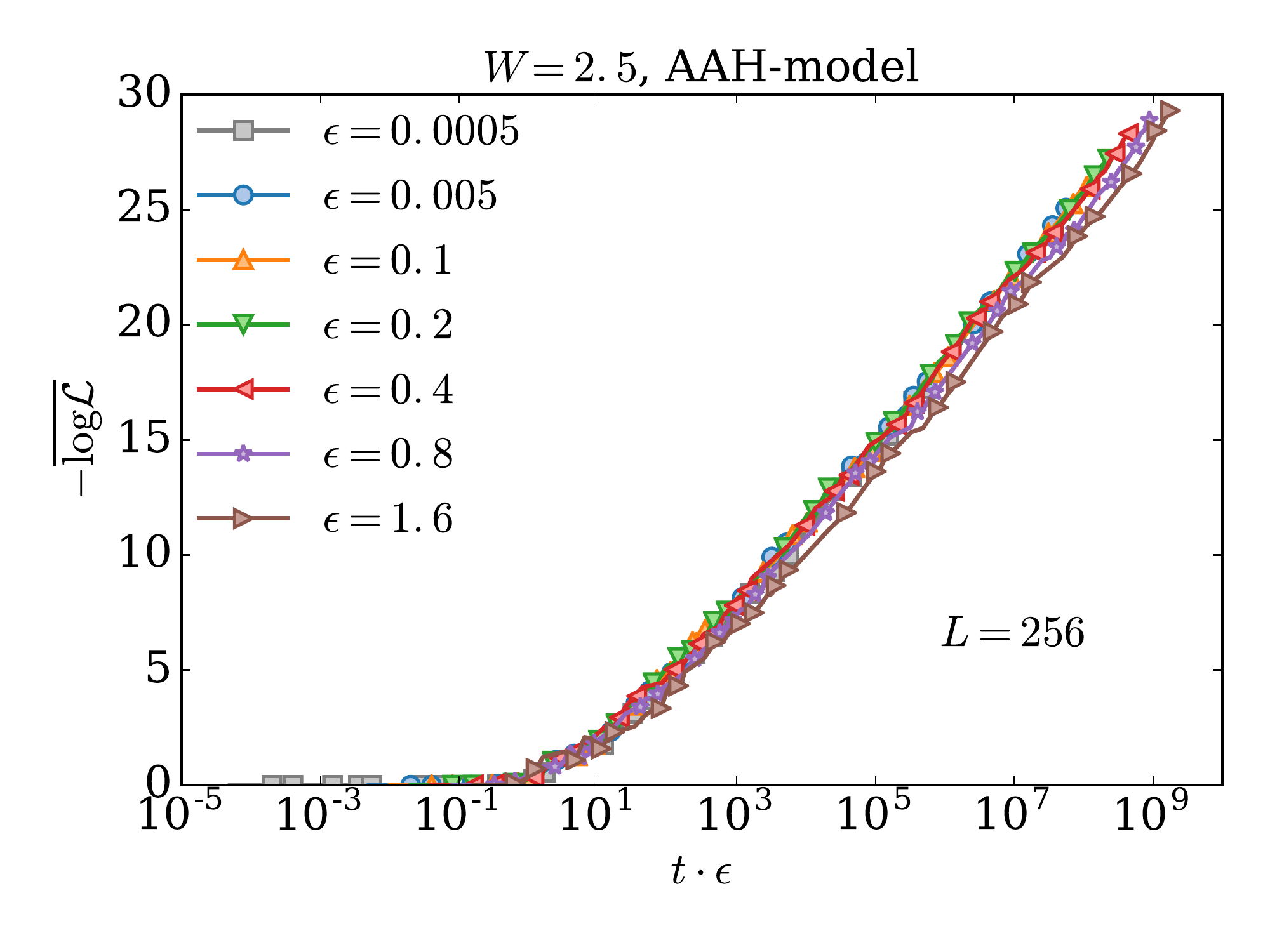}
 \caption{$-\overline{\log{\mathcal{L}}}(t)$ for different values of $\epsilon$ for the AAH-model in the localized phase $(W=2.5)$. The time as been rescaled by $\epsilon$ as is suggested by the equation in main text $|\phi_j(0)|^2t\epsilon \approx 1$.}
 \label{fig:S1_epsilon_AAH}
\end{figure}

\begin{figure}
 \includegraphics[width=1.\columnwidth]{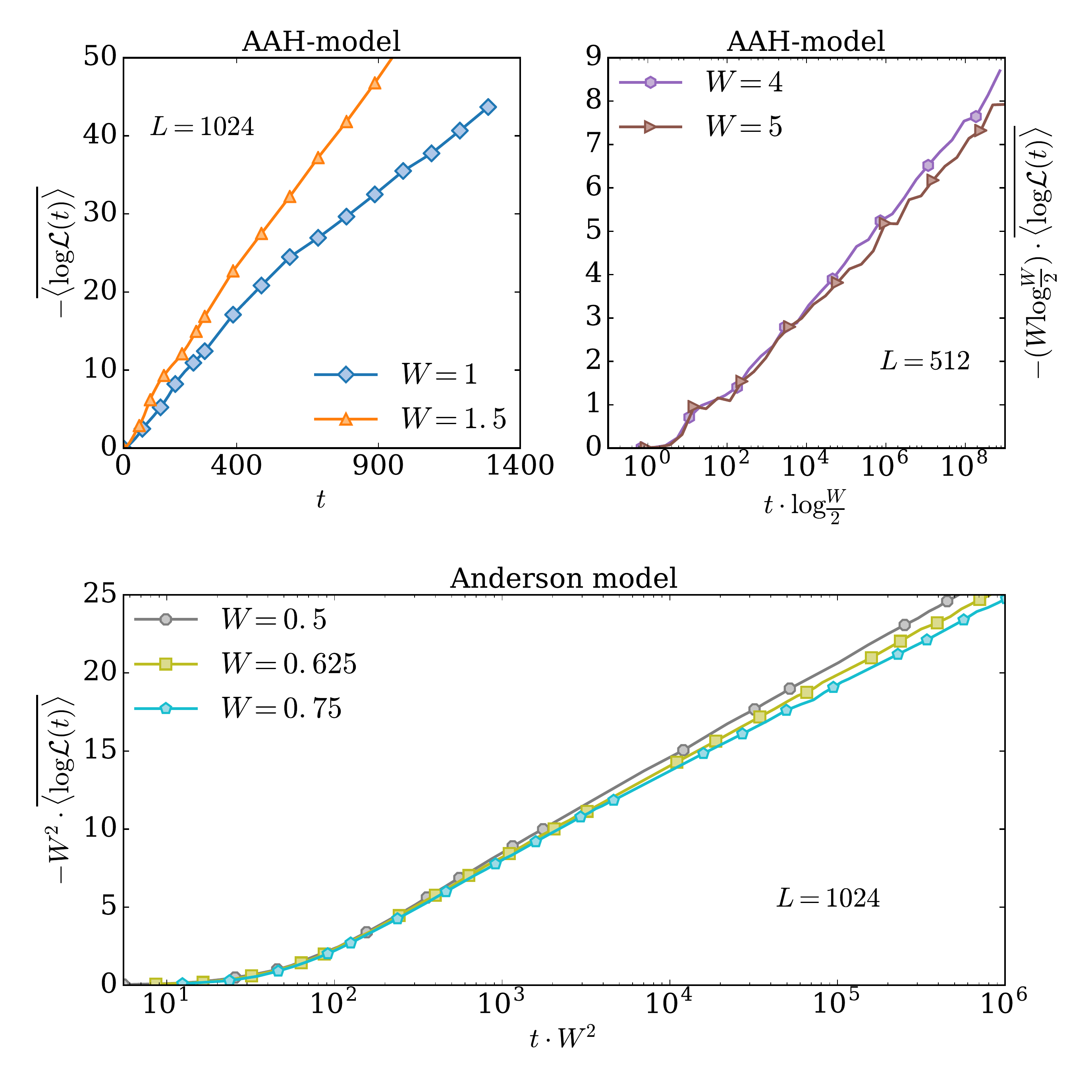}
 \caption{The panels show $-\log{\mathcal{L}}(t)$ averaged over random product states and disorder configurations for  the AAH-model and for the Anderson model. $\mathcal{L}(t)$ and $t$ has been rescaled as in the main text.}
 \label{fig:S2}
\end{figure}

\begin{figure}
 \includegraphics[width=1.\columnwidth]{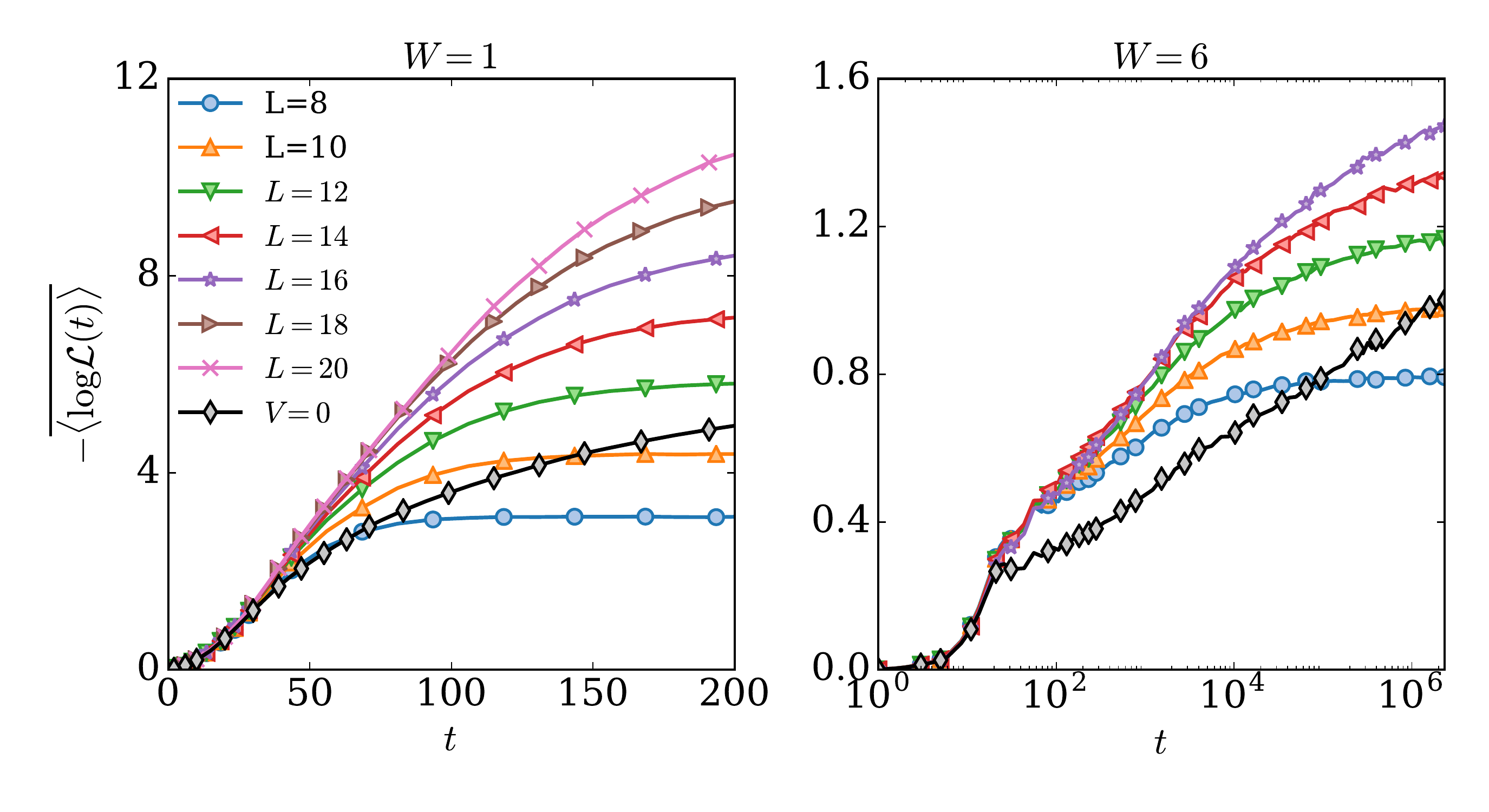}
 \caption{The panels show $-\log{\mathcal{L}}(t)$ averaged over random product states and disorder configurations for the interacting spinless t-V chain for two different $W$ ($W=1$ in the ergodic phase,$W=6$ in the localized phase).}
 \label{fig:S3}
\end{figure}

In this section, we show the dependence of $\mathcal{L}(t)$ on $\epsilon$ and the initial state $|\psi \rangle$. % over random product states of the form $\big (\prod_{s=1}^N c^\dagger_{i_s} |0\rangle \big )$.
Figure \ref{fig:S1_epsilon_AAH} shows $\mathcal{L}(t)$ for several values of the perturbation strength $\epsilon$ for the AAH-model in the localized phase $(W=2.5)$. $\mathcal{L}(t)$ decays algebraically, and, as predicted in the main text, the exponent of the 
decay $\beta$ does not depend on $\epsilon$. Moreover, as suggested by the expression $|\phi_j(0)|^2t\epsilon$, the time has been properly rescaled to make the curves collapse together.  
\begin{figure}
 \includegraphics[width=1.\columnwidth]{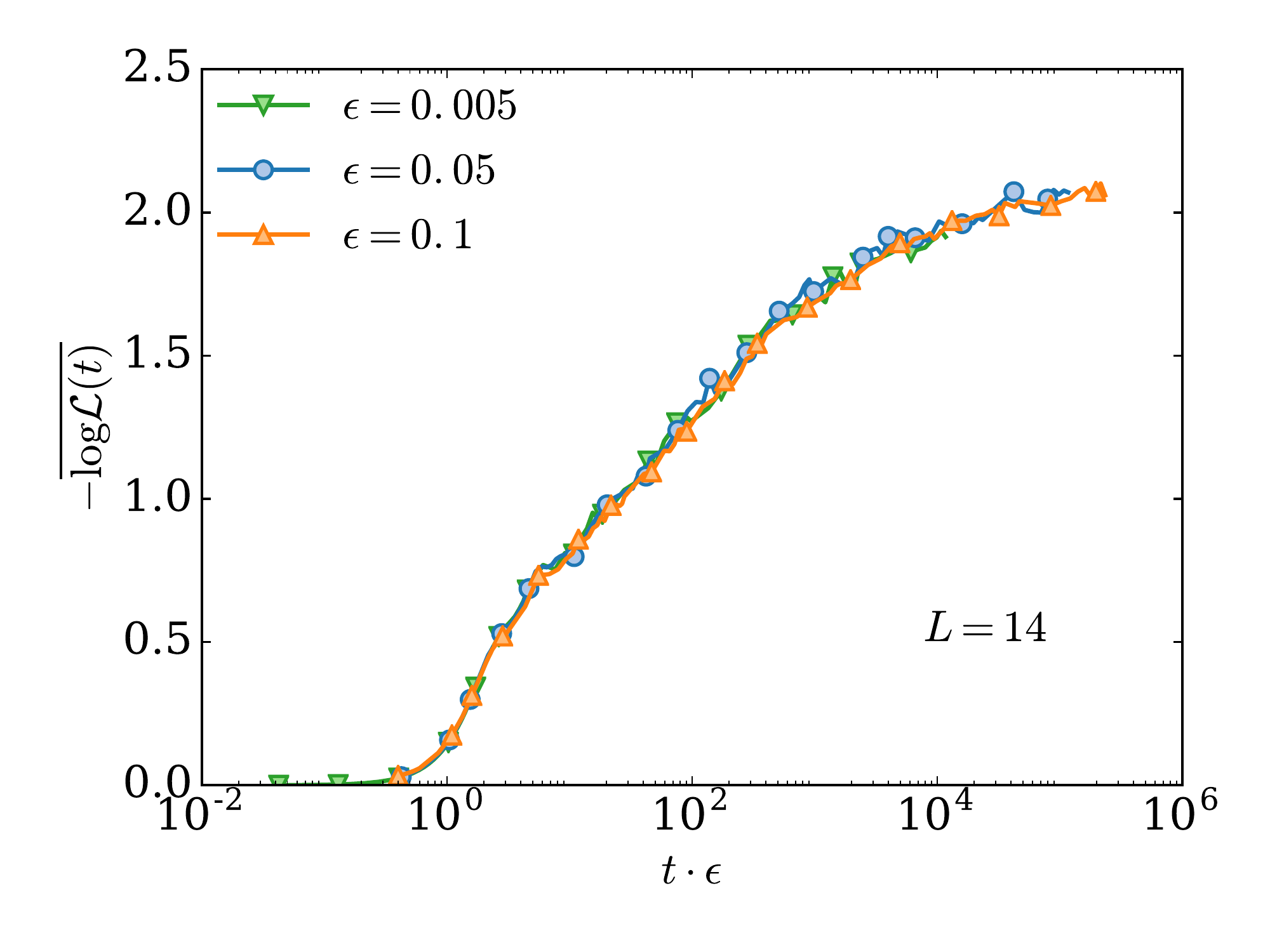}
 \caption{The panels show $-\log{\mathcal{L}}(t)$ averaged over disorder configurations for the interacting spinless t-V chain in the localized phase $(W=6)$ for a fixed system size $L=14$ for several values of the perturbation strength $\epsilon$.}
 \label{fig:S4}
\end{figure}
Figure~\ref{fig:S4} shows that the same rescaling for the time $(t\rightarrow t\epsilon$) also holds in the the interacting case.
To check that our results are qualitatively independent of the choice of the initial product state, we also averaged the LE over random product states of the form $\prod_{s=1}^N c^\dagger_{i_s} |0\rangle $ (for any random configuration we calculate $\mathcal{L}(t)$ for 25 random product states), as shown in Fig. \ref{fig:S2} (the average over 
random product states is indicated with $\langle \cdot \rangle$). Figure~\ref{fig:S2} shows that the behavior of $\mathcal{L}(t)$ is similar to the Neel state, and the same scaling with the microscopic parameter $W$ works still relatively well, and the deviations become relevant for long times ($t > 10^4$).
Figure \ref{fig:S3} shows  the behavior of LE averaged over random product states and disorder configurations for the interacting t-V chain for two different values of $W$, in the ergodic $(W=1)$  phase $\mathcal{L}(t) \sim e^{-\Gamma t}$ and in the localized $(W=6)$
phase $\mathcal{L}(t)\sim t^{-\beta}$. 
\subsection{$-\log{\overline{\mathcal{L}}}(t)$ v.s. -$\overline{\log{\mathcal{L}(t)}}$} 
\begin{figure}
 \includegraphics[width=1.\columnwidth]{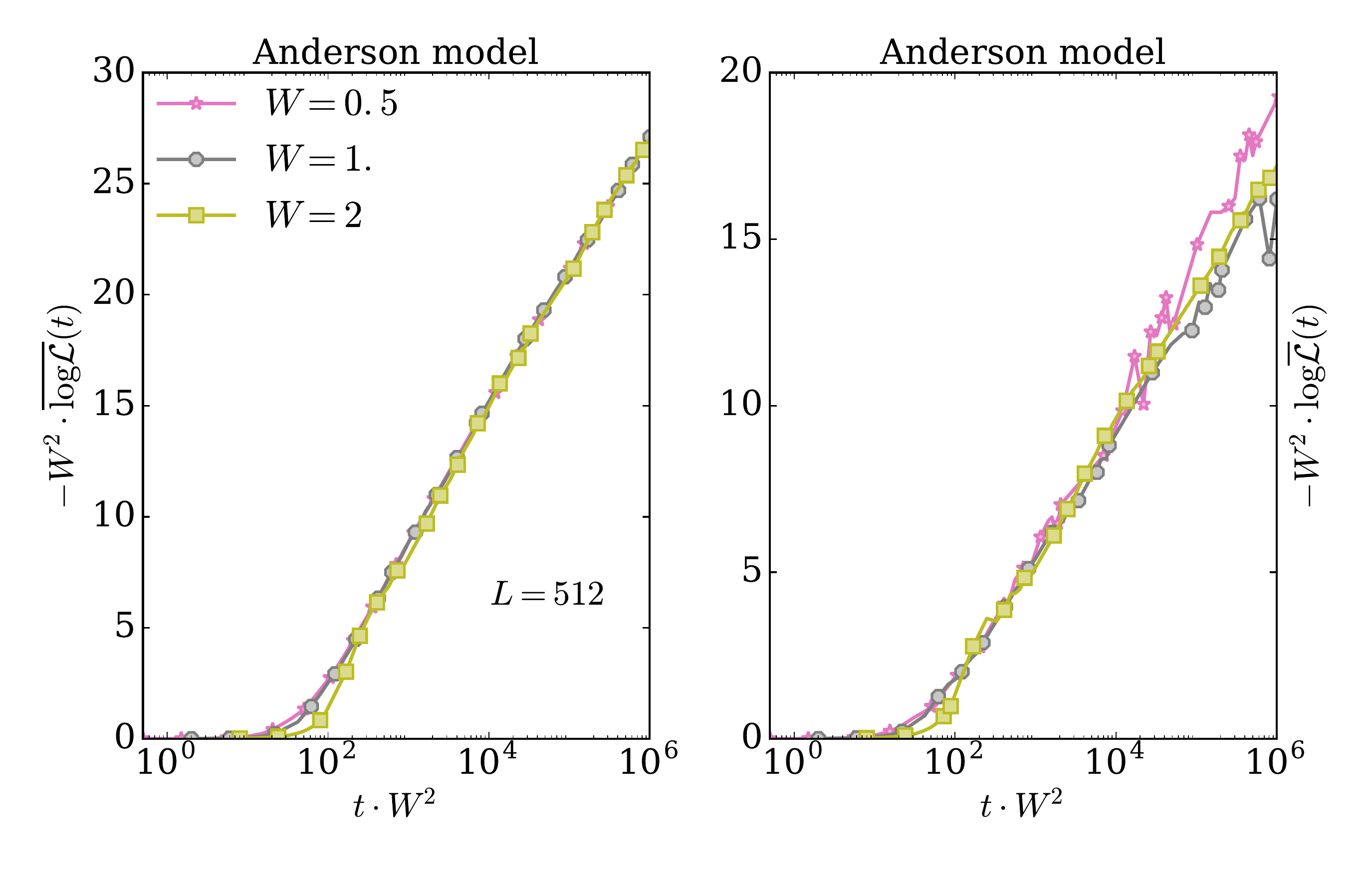}
 \caption{The panels shows $-\overline{\log{\mathcal{L}}}(t)$ (left) and $-\log{\overline{\mathcal{L}}}(t)$ (right) for the non interacting Anderson model for different values of $W$.}
 \label{fig:S5}
\end{figure}

\begin{figure}
 \includegraphics[width=1.\columnwidth]{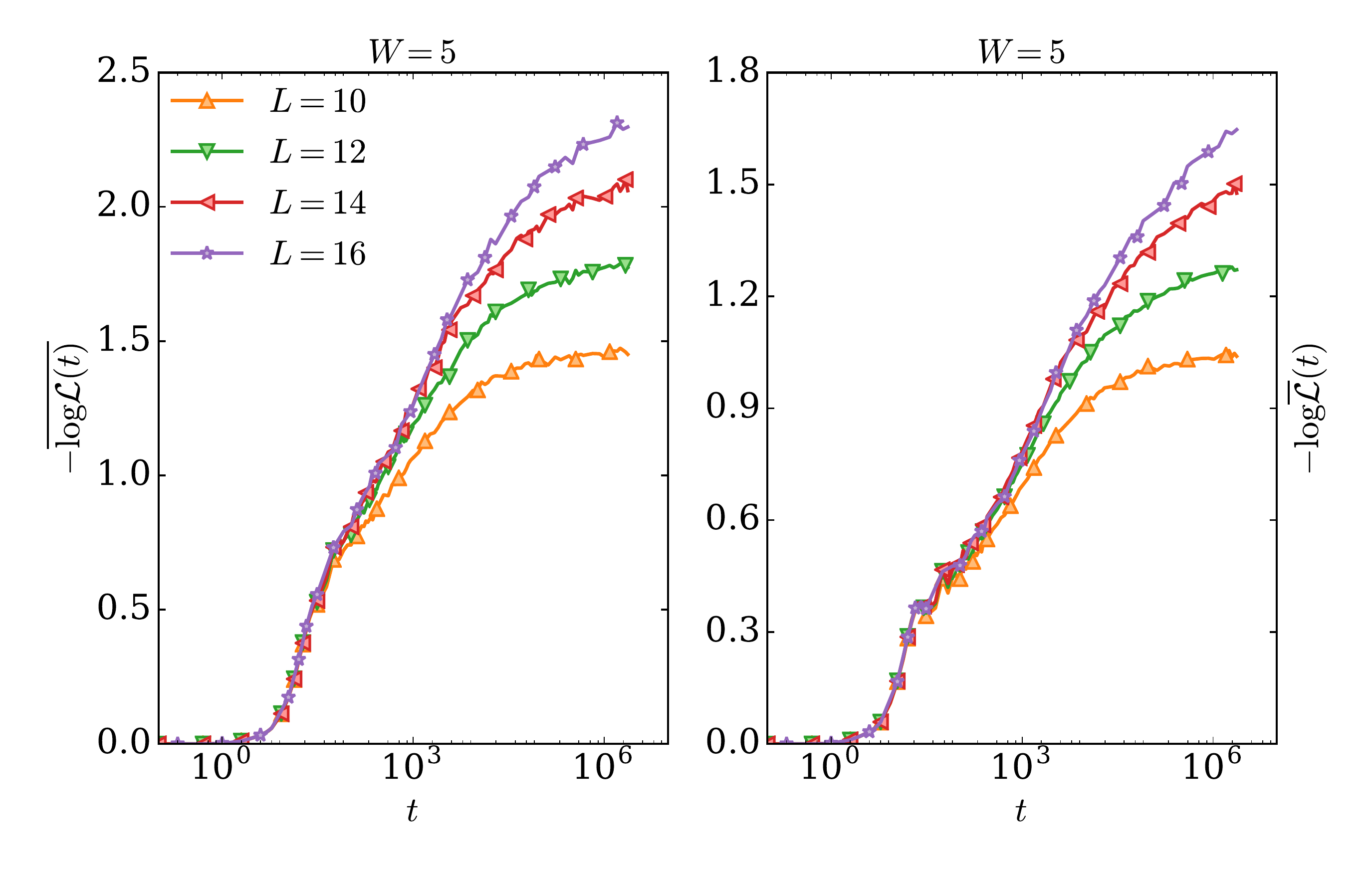}
 \caption{The panels shows $-\overline{\log{\mathcal{L}}}(t)$ (left) and $-\log{\overline{\mathcal{L}}}(t)$ (right) for the interacting t-V chain for $W=5$.}
 \label{fig:S6}
\end{figure}

Figure \ref{fig:S5} shows for the non-interacting Anderson model and for different disorder strengths $-\overline{\log{\mathcal{L}}(t)}$ and $-\log{\overline{\mathcal{L}}}(t)$. As expected from the inequality between the arithmetic mean and the geometric mean, $-\overline{\log{\mathcal{L}}(t)}\ge -\overline{\log{\mathcal{L}}(t)}$. Moreover $-\overline{\log{\mathcal{L}}(t)}$, being the logarithm of the typical value of $\mathcal{L}(t)$, is less noisy than the logarithm of the arithmetic mean  $-\log{\overline{\mathcal{L}}}(t)$.
Nevertheless, to some extent the scaling of the algebraic decay $\beta$ and of the time are the same for both ways of performing the disorder average. As shown in Fig.~\ref{fig:S6}, there is not qualitative but only quantitative difference between the typical and average values also for the interacting t-V chain. 

\subsection{$\mathcal{L}_A(t)$} 
In this section, we show the derivation of the approximate formula $\mathcal{L}_A(t)$. The main assumptions will use are the following: First, the perturbation modifies only 
the eigenenergies of $\mathcal{H}_\epsilon$  but not its eigenfunctions, which are the same as those of the unperturbed Hamiltonian $\mathcal{H}$. It is easy to see that the contribution to the change of the eigenfunctions is second order in the strength of the perturbation $\epsilon$. Second, the behavior of LE is independent of the initial choice of the product state. Using the spectral representation for the time evolution
\begin{equation}
 \mathcal{L}(t)  =  \left | \sum_{n,m}  \langle \psi | n \rangle \langle n | m_{\epsilon}\rangle \langle m_{\epsilon} |\psi \rangle e^{-it(E_n-E_m^{(\epsilon) })}  \right |^2,
\end{equation}
With the use of the first approximation  $\langle n | m_{\epsilon}\rangle = \delta_{n,m}$,
\begin{equation}
 \mathcal{L}(t)  \approx  \left | \sum_{n}  |\langle n  |\psi\rangle |^2  e^{-it(E_n-E_n^{(\epsilon) })}  \right |^2,
\end{equation}
and using the second approximation, 
\begin{equation}
 \mathcal{L}(t)  \approx  \left | \frac{1}{2^L} \sum_{n}   e^{-it(E_n-E_n^{(\epsilon) })}  \right |^2,
\end{equation}
Using first-order perturbation theory in $\epsilon$ to estimate the energy difference as  $E_n-E_n^{(\epsilon) }$, we get
\eq{
 \mathcal{L} (t) \approx & \left | \frac{  \sum_{n}    e^{-i 2 t \epsilon \langle n | \hat{n}_0 | n \rangle } }{2^L}  \right |^2
}
Moreover  $\langle n|\hat{n}_0|n \rangle = \sum_{j=1}^{L} a_j^{(n)} |\phi_j(0)|^2$,
where $\{\phi_j(0)\}$ are the single particle wave-functions evaluated in the center of the the chain and $a_j^{(n)}$ takes only two values $\{1,0\}$ depending on whether the single-particle eigenstate labeled with $j$ is occupied or 
not in the state $|n\rangle$. Finally,
\eq{
 \mathcal{L} (t) \approx & \left |\prod_{j=1}^L  \frac{ e^{-i 2 t \epsilon |\phi_j(0)|^2}+1 }{2}  \right |^2 = \prod_{j=1}^L \cos^2 \left ( \epsilon |\phi_j(0)|^2 t\right )
}
The last expression is essentially a perturbation expansion in $\epsilon$, $\mathcal{L}(t) = \mathcal{L}_A(t)+ \mathcal{O}(\epsilon^2)$.
\subsection{$\overline{\mathcal{D}}(t)$} 
In this section, we show supplemental materials for $\overline{\mathcal{D}}(t)$. In the main text we give numerical evidence that for the extended phase for the AAH model $\overline{\mathcal{D}}_\infty \sim \sqrt{L}$. Here we give an analytical argument based on a random matrix approximation which will give a lower bound for the $L$-scaling. Neglecting the time fluctuation of $\mathcal{D}(t)$ we get a lower bound of $\mathcal{D}_\infty$ (diagonal ensemble).
\begin{equation}
 \frac{1}{T} \int_0^T \mathcal{D}(s) ds \ge \frac{1}{T} \sum_x \left |\int_0^T \delta \rho(x,s) ds\right |
\end{equation}
%\begin{equation}
%\begin{split}
%|\delta \rho(x,t\rightarrow \infty)| & = \\ 
%&  	\sim \left | \sum_{\alpha} \sum_{s}^{'} %\frac{|\tilde{\phi}_\alpha %(x)|^2|\tilde{\phi}_\alpha (s)|^2}{L}  -  %\sum_{\alpha}\sum_{s}^{'} %\frac{|\tilde{\phi}_\alpha^\epsilon %(x)|^2|\tilde{\phi}_\alpha^\epsilon (s)|^2}{L} %\right | 
%\end{split}
%\end{equation}
Thus,
\begin{equation*}
 \lim_{T\rightarrow \infty} \frac{1}{T} \left |\int_0^T \delta \rho(x,s) ds\right | =
\end{equation*}
\begin{equation}
\begin{split}
&= \left | \sum_{\alpha} \sum_{s}^{'} |\phi_\alpha (x)|^2|\phi_\alpha (s)|^2  -  \sum_{\alpha}\sum_{s}^{'} |\phi_\alpha^\epsilon (x)|^2|\phi_\alpha^\epsilon (s)|^2 \right | \\
&  	\sim \left | \sum_{\alpha} \sum_{s}^{'} \frac{|\tilde{\phi}_\alpha (x)|^2|\tilde{\phi}_\alpha (s)|^2}{L^2}  -  \sum_{\alpha}\sum_{s}^{'} \frac{|\tilde{\phi}_\alpha^\epsilon (x)|^2|\tilde{\phi}_\alpha^\epsilon (s)|^2}{L^2} \right | 
\end{split}
\end{equation}
The sum over the index $s$ runs over the index sites that are occupied at the initial time ($t$=$0$). $\{\phi_\alpha(x) \}_{1}^{L}$ and $\{ \phi_\alpha^\epsilon (x) \}_{1}^{L}$ as in the main text are the single particle wave-functions of $\hat{\mathcal{H}}$ and $\hat{\mathcal{H}}_\epsilon$ respectively. In first approximation in the extended phase, the single particle wave-functions can be approximated with $\{ \frac{\tilde{\phi}_\alpha(x)}{\sqrt{L}} \}_{1}^{L}$ and $\{ \frac{\tilde{\phi}_\alpha^\epsilon (x)}{\sqrt{L}} \}_{1}^{L}$, where $\{ \tilde{\phi}_\alpha(x) \}_{1}^{L}$ and $\{ \tilde{\phi}_\alpha^\epsilon (x) \}_{1}^{L}$ are independent random variables with a fixed mean and variance which does not scale with $L$, since its dependence on $L$ has been already taken care with the normalization factor $\frac{1}{\sqrt{L}}$. Using the central limit theorem, we can estimate the scaling with $L$ of the sum over the index $s$ and $\alpha$, e.g.
$\sum_{s}^{'} \frac{|\tilde{\phi}_\alpha (s)|^2}{L}\sim constant_1 + \mathcal{O}(\frac{1}{\sqrt{L}})$ and $\sum_{\alpha} \frac{|\tilde{\phi}_\alpha (x)|^2}{L}\sim constant_2 + \mathcal{O}(\frac{1}{\sqrt{L}})$. Since we have assumed that the perturbation does not change the statistical properties of the single particle wave-functions, we have that the difference of the local density profile $|\rho(x, t\rightarrow \infty)| \ge \mathcal{O}(\frac{1}{\sqrt{L}})$. This gives the result $\overline{\mathcal{D}}_\infty \ge \mathcal{O}(\sqrt{L})$.
The argument can be repeated for the non-interacting localized phase giving a lower bound $\overline{\mathcal{D}}_\infty\ge \mathcal{O}(L)$. The difference is that now the single particle wave-functions should be taken in first approximation as box functions with a finite width randomly displaced, (e.g. $\phi_\alpha (x) \sim \frac{\chi_{[\alpha -\xi,\alpha+\xi]}}{\sqrt{2\xi}}$). 
\begin{figure}
 \includegraphics[width=1.\columnwidth]{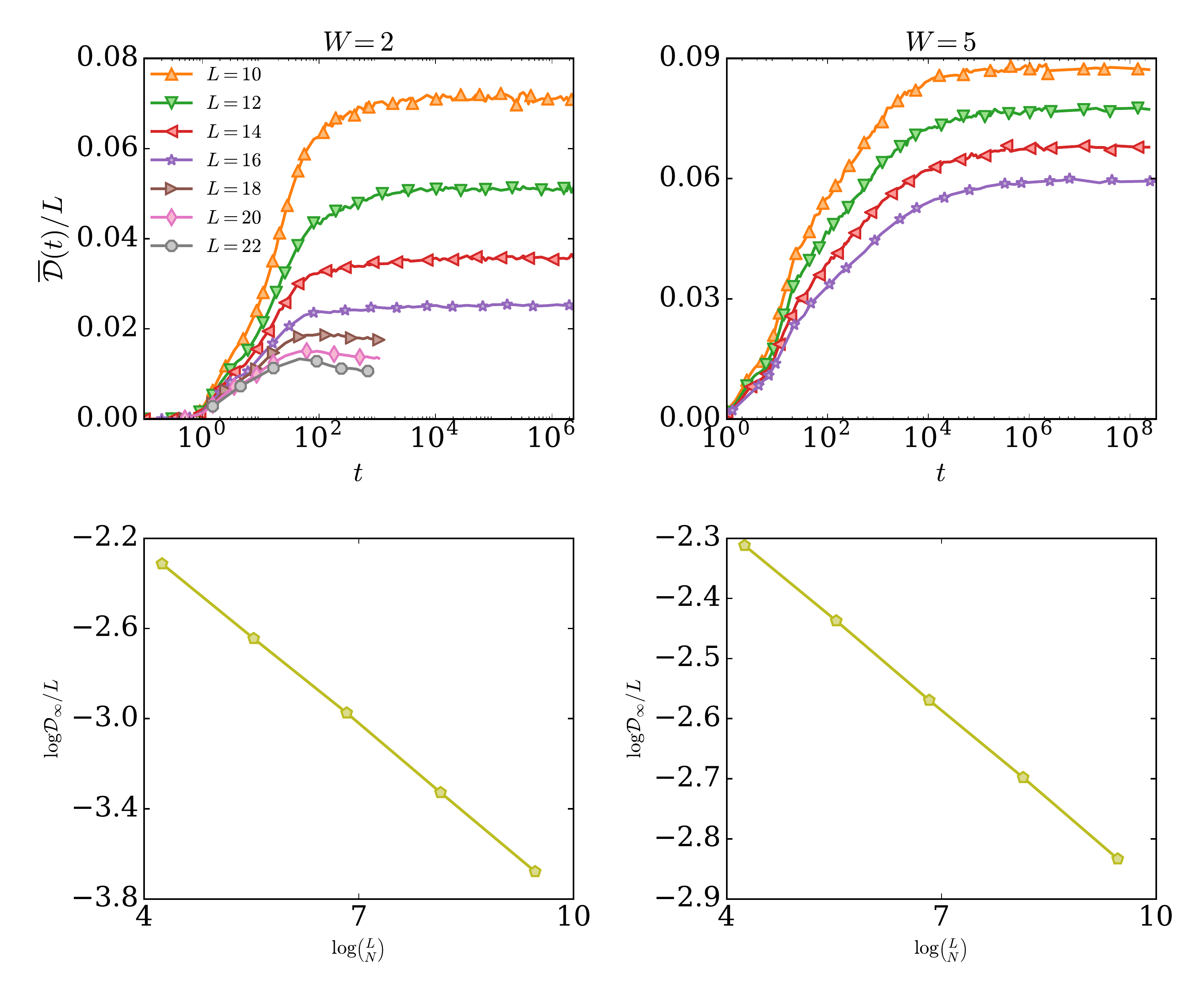}
 \caption{The top panel shows $\frac{\overline{D}(t)}{L}$ for $W=2$ (ergodic) and $W=5$ (localized) for the interacting t-V chain for several system sizes $L$. The bottom panels show that in both phases $\overline{\frac{\mathcal{D}_\infty}{L}} \sim  \binom{L}{N}^{-\gamma}$.}
 \label{fig:S7}
\end{figure}

In the following, we show some supplemental material of $\overline{\mathcal{D}}(t)$ for the interacting problem. Figure~\ref{fig:S7} shows the behavior of $\overline{\mathcal{D}}(t)$ for different system sizes in the two phases of the t-V chain. For $W=2$, the system is in the ergodic phase, and $\overline{\mathcal{D}}(t)$ exhibits the same non-monotonic behaviour as a function of $t$ as for the case $W=1$ (shown in the main text). For $\overline{\mathcal{D}}(t)$ in the localized phase $(W=5)$, finite size effects are important. Indeed, it is not possible to see the non-monotonic phase even in the ergodic phase for system sizes smaller than $L\leq 16$. However, Fig.~\ref{fig:S7} gives evidence that in both
phases, $\overline{\mathcal{D}}_\infty \sim L \binom{L}{n}^{-\gamma}$. In the ergodic phase for $W=2$, $\gamma \approx 0.26$. In the localized phase for $W=5$, 
the exponent $\gamma \approx 0.1$. The exponent is small so that for system size $L\le 16$ the behavior of the function $L \binom{L}{N}^{-\gamma}$ is dominated by the linear part $L$. Nevertheless, if this scaling persists in the thermodynamic limit, $\overline{\mathcal{D}}(t)$ in the 
long-time limit will go to zero.
\end{document}